\documentclass{article}
\usepackage{spconf,amsmath,graphicx}
\usepackage{times}
\usepackage{graphicx}
\usepackage{amsmath}
\usepackage{amssymb}
\usepackage{algorithm}
\usepackage{tikz}
\usepackage{subfig}
\usetikzlibrary{spy,calc}
\usepackage{algorithmic}
\usepackage{enumerate}
\usepackage{multirow}
\usepackage{multicol}
\usepackage[font=footnotesize,labelformat=simple]{subcaption}
\usepackage{arydshln}
\usepackage{tikz}
\usetikzlibrary{spy}
\usepackage[hyphens,spaces,obeyspaces]{url}
\usepackage[colorlinks,allcolors=blue]{hyperref}
\newif\ifblackandwhitecycle

\usepackage{color}
\usepackage{amssymb}
\usepackage{bigstrut}
\def\0{{\mathbf 0}}
\def\1{{\mathbf 1}}

\def\e{{\mathbf e}}
\def\f{{\mathbf f}}

\def\u{{\mathbf u}}
\def\v{{\mathbf v}}

\def\x{{\mathbf x}}
\def\y{{\mathbf y}}
\def\z{{\mathbf z}}

\def\B{{\mathbf B}}
\def\C{{\mathbf C}}
\def\D{{\mathbf D}}

\def\I{{\mathbf I}}

\def\L{{\mathbf L}}
\def\M{{\mathbf M}}

\def\V{{\mathbf V}}
\def\W{{\mathbf W}}

\def\ie{{\textit{i.e.}}}

\def\cE{{\mathcal E}}

\def\cG{{\mathcal G}}

\def\cS{{\mathcal S}}

\def\cV{{\mathcal V}}

\def\bxi{{\boldsymbol \xi}}
\def\bLambda{{\boldsymbol \Lambda}}

\newcommand{\comment}[1]{}

\title{Unrolling Nonconvex Graph Total Variation for Image Denoising}
\name{{Songlin Wei$^\star$, Gene Cheung$^\dag$, Fei Chen$^\star$, Ivan Selesnick$^\#$}
\thanks{The work of G. Cheung was supported in part by the Natural Sciences and Engineering Research Council of Canada (NSERC) RGPIN-2025-06252. The work of F. Chen was supported in part by the National Natural Science Foundation of China (62471141).}}
\address{$^\dag$York University, Canada ~~~~~~ $^\star$Fuzhou University, China ~~~~~~ $^\#$New York University, USA }
\begin{document}
%
\maketitle
\begin{abstract}
Conventional model-based image denoising optimizations employ convex regularization terms, such as total variation (TV) that convexifies the $\ell_0$-norm to promote sparse signal representation.
Instead, we propose a new non-convex total variation term in a graph setting (NC-GTV), such that when combined with an $\ell_2$-norm fidelity term for denoising, leads to a convex objective with no extraneous local minima.
We define NC-GTV using a new graph variant of the Huber function, interpretable as a Moreau envelope.
The crux is the selection of a parameter $a$ characterizing the graph Huber function that ensures overall objective convexity; 
we efficiently compute $a$ via an adaptation of Gershgorin Circle Theorem (GCT).
To minimize the convex objective, we design a linear-time algorithm based on Alternating Direction Method of Multipliers (ADMM) and unroll it into a lightweight feed-forward network for data-driven parameter learning.
Experiments show that our method outperforms unrolled GTV and other representative image denoising schemes, while employing far fewer network parameters. 
\end{abstract}
\section{Introduction}
\label{sec:intro}
Image denoising \cite{milanfar13} is a well-studied basic restoration problem that remains important today; other image restoration tasks like deblurring \cite{bai19} can be addressed using general frameworks like the \textit{Plug-and-Play} \textit{Alternating Direction Method of Multipliers} (PnP-ADMM) \cite{chan17} that employ an image denoiser as a building block, and popular generative diffusion models can be implemented as cascades of image denoisers \cite{chan24}.
While \textit{deep learning} (DL) based denoisers \cite{DnCNN} are now prevalent and achieve state-of-the-art (SOTA) performance, they are inherent ``black boxes" that are uninterpretable and employ a huge number of network parameters, and thus are not suitable for memory-constrained devices.
In contrast, an alternative hybrid model-based / data-driven paradigm called \textit{algorithm unrolling} \cite{monga21} builds an interpretable neural net by implementing each iteration of a model-based iterative algorithm as a neural layer towards a feed-forward network amenable to parameter learning.
We pursue this paradigm here for image denoising, but from a \textit{graph signal processing} (GSP) perspective \cite{ortega18ieee,cheung18}. 

Model-based image denoisers commonly formulate an optimization with a convex regularization term such as \textit{total variation} (TV) \cite{strong03}---convexification of the combinatorial $\ell_0$-norm to promote sparse signal restoration (\ie, sparse discontinuities for an otherwise smooth signal). 
However, a large signal discontinuity $|x_{i+1} - x_i|$ at pixel $i$ has much larger $\ell_1$-norm than $\ell_0$-norm, and thus minimizing TV tends to underestimate signal discontinuities, resulting in sub-optimal reconstructions of \textit{piecewise constant} (PWC) signals.

To improve upon TV, for one-dimensional signals \cite{Selesnick2020} pursued a \textit{convex non-convex} (CNC) strategy \cite{selesnick15,lanza2019sparsity}: combine a non-convex regularization term (called the \textit{minimax-concave} (MC) function)---which reduces the underestimation of signal discontinuities---with a convex fidelity term, so that the overall objective remains convex, thereby mitigating the problem of extraneous local minima.
\cite{Selesnick2020} demonstrated better PWC signal recovery than TV for 1D signal denoising.

Inspired by \cite{Selesnick2020}, we propose \textit{non-convex graph total variation} (NC-GTV)---extension of MC function to the graph domain---for image denoising. 
Specifically, we define NC-GTV using a new graph variant of the \textit{Huber function} \cite{huber64}, which is interpretable as a Moreau envelope \cite{heinz17}.
The crux is the selection of a MC parameter $a$ characterizing the graph Huber function that ensures overall objective convexity; 
we efficiently compute $a$ via a clever adaptation of \textit{Gershgorin Circle Theorem} (GCT) \cite{varga04}.
To minimize the convex objective, we design a linear-time algorithm based on \textit{Alternating Direction Method of Multipliers} (ADMM) \cite{parikh13} and unroll it into a lightweight feed-forward network for data-driven parameter learning.
Experiments show that our method outperforms unrolled GTV and other representative model-based and deep-learning-based image denoising schemes \cite{dabov07,DnCNN}, while employing drastically fewer network parameters. 


We summarize our contributions as follows:
\begin{enumerate}
\item We design a graph Huber function $S_a(\x',\x)$---extension of the scalar Huber function $s_a(x)$ to the graph domain---as a quadratic term using penalty graph Laplacian $\L_a^p$ that is amenable to fast optimization.
Like $s_a(x)$, we show $S_a(\x',\x)$ is interpretable as a Moreau envelope.
\item Leveraging GCT, we efficiently compute the ``tightest" parameter $a$ characterizing the graph Huber function that guarantees PSDness of coefficient matrix $\I - \mu \L_a^p$, and hence convexity of the optimization objective. 
\item We unroll an ADMM algorithm into a lightweight interpretable feed-foward network for data-driven parameter learning, with competitive denoising performance.
\end{enumerate}



\section{Preliminaries}
\label{sec:prelim}
\subsection{GSP Definitions}

A positive undirected graph $\cG = (\cV, \cE, \W)$ is defined by a node set $\cV = \{1, \ldots, N\}$ and an edge set $\cE = \{(i,j)\}$ with $M = |\cE|$ edges, where edge $(i,j) \in \cE$ has non-negative weight $w_{i,j} = W_{i,j}$, for \textit{adjacency matrix} $\W \in \mathbb{R}^{N \times N}$. 
Undirected edges mean that $W_{i,j} = W_{j,i}$, and $\W$ is symmetric.
\textit{Degree matrix} $\D \in \mathbb{R}^{N \times N}$ is a diagonal matrix with diagonal entries $D_{i,i} = \sum_j W_{i,j}$. 
The combinatorial \textit{graph Laplacian matrix} $\L \in \mathbb{R}^{N \times N}$ \cite{ortega18ieee} is defined as
\begin{align}
\L \triangleq \D - \W = \mathrm{diag}(\W \1) - \W,
\end{align}
where $\1$ is an all-one vector of appropriate length, and $\mathrm{diag}(\v)$ is the diagonal matrix containing vector $\v$ as diagonal entries.
$\L$ is provably \textit{positive semi-definite} (PSD) if all edge weights are non-negative, \ie, $W_{i,j} \geq 0, \forall i,j$ \cite{cheung18}. 

Real and symmetric Laplacian $\L$ can be eigen-decomposed to $\L = \V \bLambda \V^\top$, where $\V$ contains eigenvectors of $\L$ as columns, and $\bLambda = \mathrm{diag}(\lambda_1, \ldots, \lambda_N)$ is a diagonal matrix with ordered eigenvalues $0 = \lambda_1 \leq \lambda_2 \leq \ldots \leq \lambda_N$ along its diagonal.
The $k$-th eigen-pair $(\lambda_k,\v_k)$ is the $k$-th graph frequency and Fourier mode for $\cG$, respectively.
$\tilde{\x} = \V^\top \x$ is the \textit{graph Fourier transform} (GFT) of signal $\x$ \cite{ortega18ieee}, where $\tilde{x}_k = \v_k^\top \x$ is the $k$-th GFT coefficient for signal $\x$.

\subsection{Graph Laplacian Regularizer}

To regularize an ill-posed graph signal restoration problem like denoising \cite{pang17} or quantization \cite{liu17}, \textit{graph Laplacian regularizer} (GLR) is popular due to its convenient quadratic form that is amenable to fast optimization \cite{pang17}. 
Given a positive graph $\cG$ specified by graph Laplacian $\L$, GLR for a signal $\x \in \mathbb{R}^N$ is defined as
\begin{align}
\x^\top \L \x &= \sum_{(i,j) \in \cE} w_{i,j} (x_i - x_j)^2 = \sum_k \lambda_k \tilde{x}_k^2 .
\label{eq:GLR} 
\end{align}
Thus, minimizing GLR means promoting a signal with connected sample pairs $(x_i,x_j)$ that are similar.
In the spectral domain, it means promoting a \textit{low-pass} signal with energies $\tilde{x}_k^2$'s concentrated in low graph frequencies $\lambda_k$'s. 


\subsection{Graph Total Variation}

Another popular graph signal regularization term is \textit{graph total variation} (GTV) \cite{elmoataz08,couprie13,bai19}. 
Define first the \textit{incidence matrix} $\C \in \mathbf{R}^{M \times N}$ as

\vspace{-0.1in}
\begin{small}
\begin{align}
C_{m,j} = \left\{ \begin{array}{ll}
w_{i,j} & \mbox{if start node of $m$-th edge $(j,i)$ is $j$} \\
-w_{i,j} & \mbox{if end node of $m$-th edge $(i,j)$ is $j$} \\
0 & \mbox{o.w.}
\end{array} \right. .
\end{align}
\end{small}\noindent
Given an undirected graph $\cG$, the start and end nodes of each edge $(i,j)$ are arbitrary. 
We can now define GTV as
\begin{align}
\|\C \x\|_1 = \sum_{(i,j) \in \cE} w_{i,j} |x_i - x_j| .
\label{eq:GTV}
\end{align}
\eqref{eq:GTV} is essentially the $\ell_1$-norm variant of GLR \eqref{eq:GLR}.

\section{Problem Formulation}
\label{sec:formulate}
\subsection{Defining Graph Huber Function}

\subsubsection{MC penalty and Huber Functions}

From \cite{Selesnick2020}, the scalar \textit{minimax-concave} (MC) penalty function $\phi_a(x): \mathbb{R} \mapsto \mathbb{R}_+$ is defined as
\begin{align}
\phi_a(x) = \left\{ \begin{array}{ll}
|x| - \frac{a}{2} x^2 & \mbox{if}~ |x| \leq \frac{1}{a} \\
\frac{1}{2a} & \mbox{o.w.}
\end{array} \right. 
\label{eq:MC}
\end{align}
where $a > 0$ is a non-negative \textit{MC parameter}. 
Note that for $a=0$, $\phi_a(x) = |x|$, \ie, the $\ell_1$-norm.

The MC penalty function can be written in terms of the \textit{Huber function} \cite{huber64} $s_a(x)$, \ie, $\phi_a(x) = |x| - s_a(x)$, where $s_a(x)$ is defined as 
\begin{align}
s_a(x) = \left\{ \begin{array}{ll}
\frac{a}{2} x^2 & \mbox{if}~ |x| \leq \frac{1}{a} \\
|x| - \frac{1}{2a} & \mbox{o.w.}
\end{array} \right. .
\label{eq:Huber}
\end{align}
The Huber function $s_a(x)$ can also be written in terms of the \textit{Moreau envelope} \cite{heinz17} $f^M(x)$ defined as
\begin{align}
f^M(\x) &\triangleq \inf_{\v \in \mathbb{R}^N} \left\{
f(\v) + \frac{1}{2} \|\x - \v\|^2_2 \right\} 
\label{eq:moreau} \\
s_a(x) &= \min_{v} \left\{
|v| + \frac{a}{2}(x - v)^2 
\right\} = a (\frac{1}{a} |\cdot|)^M(x)
.
\end{align}

\subsubsection{Graph Huber Function}

We construct a graph variant of $s_a(x)$ for graph signal $\x \in \mathbb{R}^N$, called \textit{graph Huber function}, given a positive undirected graph $\cG(\cV,\cE,\W)$.
The idea is to mimic \eqref{eq:Huber}, where for each connected pair $(i,j) \in \cE$, we construct an $\ell_2$-norm $w_{i,j}(x_i - x_j)^2$ scaled by $\frac{a}{2}$ if $|x_i - x_j| \leq \frac{1}{a}$, and an $\ell_1$-norm $w_{i,j}|x_i - x_j|$ minus constant $\frac{w_{i,j}}{2a}$ otherwise.
We seek to express the sum of terms for all connected pairs $(i,j)$ in a compact quadratic form $\x^\top \L_a^p \x$, for some graph Laplacian matrix $\L_a^p$ ($p$ means ``penalty" here, while $a$ parameterizes $\L_a^p$).

For each edge $(i,j) \in \cE$, we first define a \textit{signal-dependent} edge weight $w^p_{i,j}$ for a corresponding \textit{penalty graph} $\cG^p$ as

\vspace{-0.05in}
\begin{small}
\begin{align}
w_{i,j}^p = \left\{ \begin{array}{ll}
\frac{a}{2} w_{i,j} & \mbox{if}~ |x_i' - x_j'| \leq \frac{1}{a} \\
\frac{w_{i,j}}{\max (|x_i' - x_j'|, \epsilon)} - \frac{w_{i,j}}{2a \max ((x_i' - x_j')^2, \epsilon)} & \mbox{o.w.}
\end{array} \right. 
\label{eq:edgeWeightP}
\end{align}
\end{small}\noindent
where $x_i'$ is the most recent estimate of signal sample at node $i$, and $\epsilon > 0$ is a small positive parameter for numerical stability, similarly done in \cite{bai19}. 

The set of edge weights $\{w^p_{i,j}\}$ thus specifies an adjacency matrix $\W^p_a$, and we can define a corresponding combinatorial graph Laplacian as $\L^p_a \triangleq \mathrm{diag}(\W^p_a \1) - \W^p_a$.
Thus, signal-dependent GLR $\x^\top \L^p_a(\x') \x$ is
\begin{align}
&\approx \!\!\!\!\!\!\! \sum_{(i,j) \in \cE \,|\, |x_i' - x_j'| \leq \frac{1}{a}} \!\!\!\!\!\!\!\!\!\!\! \frac{a}{2} w_{i,j} (x_i - x_j)^2 + \!\!\!\!\!\!\!
\sum_{(i,j) \in \cE \,|\, |x_i' - x_j'| > \frac{1}{a}} \!\!\!\!\!\!\!\!\!\!\! w_{i,j} |x_i - x_j|
\nonumber \\
& ~~~~~~ - \sum_{(i,j) \in \cE \,|\, |x_i' - x_j'| > \frac{1}{a}} \frac{w_{i,j}}{2a}
\label{eq:Huber_GLR}
\end{align}
assuming $x_i' \approx x_i, \forall i$. 
We see that \eqref{eq:Huber_GLR} is a graph variant of the Huber function \eqref{eq:Huber}: for each edge $(i,j) \in \cE$ where $|x'_i - x'_j| \leq \frac{1}{a}$, GLR computes 
$\frac{a}{2} w_{i,j} (x_i - x_j)^2$, 
and for each edge $(i,j) \in \cE$ where $|x'_i - x'_j| > \frac{1}{a}$, GLR computes 
\begin{align*}
& ~~ \frac{w_{i,j}}{|x'_i - x'_j|} (x_i - x_j)^2 - \frac{w_{i,j}}{2a(x_i'-x_j')^2}(x_i - x_j)^2 
\nonumber \\
\approx & ~~ w_{i,j} |x_i - x_j| - \frac{w_{i,j}}{2a}, 
\end{align*}
assuming $|x'_i - x'_j| > \epsilon$.
For $|x'_i - x'_j| < \epsilon$, the GLR term for edge $(i,j)$ is near zero anyway.

We can now define a signal-dependent graph Huber function $S_a(\x',\x)$ as
\begin{align}
S_a(\x',\x) \triangleq \x^\top \L^p_a(\x') \x .   
\label{eq:graph_Huber}
\end{align}
Note that the graph Huber function $S_a(\x',\x)$ is continuous w.r.t. $|x_i - x_j|$; when $|x_i - x_j| = \frac{1}{a}$, contribution from edge $(i,j)$ is $\frac{a}{2} w_{i,j} (x_i - x_j)^2 = \frac{w_{i,j}}{2a}$, and when $|x_i - x_j| = \frac{1}{a} + \gamma$, for $\gamma \rightarrow 0^+$, contribution is also $w_{i,j} |x_i - x_j| - \frac{w_{i,j}}{2a} = \frac{w_{i,j}}{2a}$.

Note also that like $s_a(x)$ in \eqref{eq:moreau}, $S_a(\x',\x)$ can also be written as a Moreau Envelope: 

\vspace{-0.1in}
\begin{small}
\begin{align}
S_a(\x',\x) \!\! = \!\!\! \min_{\v \in \mathbb{R}^M} 
\left\{
\|\v\|_1 + \frac{a}{2} \|\C\x - \v\|^2_2 
\right\}
= a (\frac{1}{a} \|\cdot \|_1)^M(\C\x) .
\end{align}
\end{small}

\subsection{Defining Objective}

We now define our signal denoising objective as follows.
The objective is a sum of the $\ell_2$-norm fidelity term, plus the graph variant of the MC penalty term \eqref{eq:MC}---GTV \eqref{eq:GTV} minus the graph Huber function \eqref{eq:graph_Huber}: 
\begin{align}
\min_\x \|\y - \x\|^2_2 + \mu \, \|\C \x\|_1 - \mu \, \x^\top \L^p_a(\x') \x ,
\label{eq:obj}
\end{align}
where $\mu > 0$ is a regularizer weight parameter. 
We observe that \eqref{eq:obj} is a $\ell_1$- / $\ell_2$-norm minimization problem. 
For the objective to be convex, we first rewrite it as 
\begin{align*}
\y^\top \y - 2 \x^\top \y + \x^\top (\I - \mu \L^p_a(\x')) \x + \mu \|\C \x\|_1
\end{align*}
which is convex if $\I - \mu \L^p_a(\x')$ is PSD.

\subsection{MC Parameter Selection}

Edge weight $w_{i,j}^p$ is computed differently in \eqref{eq:edgeWeightP} depending on $|x_i' - x_j'|$ relative to $\frac{1}{a}$.
Specifically, when $|x_i' - x_j'| = \frac{1}{a}$, edge weight $w_{i,j}^p \leftarrow \frac{a}{2} w_{i,j}$ in both computations. 
When $|x_i' - x_j'| < \frac{1}{a}$, $w_{i,j}^p$ increases linearly with $a$, and when $|x_i' - x_j'| > \frac{1}{a}$, $w_{i,j}^p$ increases more slowly with $\frac{1}{a}$.
Given that $w_{i,j}^p$ increases with $a$ generally but at two different rates, we compute a largest parameter $a^*$ while ensuring $\I - \mu \L^p_a(\x')$ is PSD, leveraging the \textit{Gershgorin Circle Theorem} (GCT) \cite{varga04}, as follows.

Define $\B \triangleq \I - \mu \L_a^p(\x')$. 
By GCT, a lower bound $\lambda_{\min}^-$ for the smallest eigenvalue $\lambda_{\min}$ of $\B$ is
\begin{align}
\lambda_{\min}^-(\B) \triangleq \min_i B_{i,i} - \sum_{j \,|\, j \neq i} |B_{i,j}| ,
\end{align}
where $c_i = B_{i,i}$ and $r_i = \sum_{j \neq i} |B_{i,j}|$ are the center and radius of the $i$-th Gershgorin disc of matrix $\B$, respectively. 
We select the largest $a$ (to maximize the effectiveness of the MC penalty function) while ensuring $\B$ is PSD, \ie,
\begin{align}
a^* = \arg \max_a ~ a, 
~~~\mbox{s.t.}~~ \lambda^-_{\min}(\B) \geq 0 .
\label{eq:max_a}
\end{align}

Given $\x'$, we sort $\frac{1}{|x_i' - x_j'|}$ for each $(i,j) \in \cE$ into a list $\cS$ of increasing magnitude; these are the $a$ values at which an edge weight $w_{i,j}^p$ switches from one computation to another via \eqref{eq:edgeWeightP}.
We identify the largest value $a^l \in \cS$ such that $\lambda_{\min}^-(\B) \geq 0$ via \textit{binary search} \cite{clr}.  
Denote by $a^u$ the next larger value in $\cS$.
We then compute the following three subset sums for each row $i$ of $\B$:
\begin{align}
\tilde{r}^{(1)}_i &= \frac{1}{2} \sum_{j|(i,j) \in \cE,\, |x'_i-x'_j|\leq \frac{1}{a^l}} w_{i,j}   
\label{eq:subset1} \\
r^{(2)}_i &= \sum_{j|(i,j) \in \cE,\, |x'_i-x'_j| > \frac{1}{a^l}} \frac{w_{i,j}}{\max(|x_i'-x_j'|,\epsilon)} 
\label{eq:subset2} \\
\tilde{r}^{(3)}_i &= \frac{1}{2} \sum_{j|(i,j) \in \cE,\, |x'_i-x'_j| > \frac{1}{a^l}} \frac{-w_{i,j}}{\max((x_i'-x_j')^2,\epsilon)} 
\label{eq:subset3}
\end{align}

Note that the $i$-th disc radius of $\B$ is $r_i = \mu (a \tilde{r}_i^{(1)} + r_i^{(2)} + \tilde{r}_i^{(3)} / a)$. 
To identify the largest $a^*_i$ for this row, we set $c_i - r_i = 0$, where $c_i = B_{i,i} = 1 - \mu (\L_a^p)_{i,i} = 1 - r_i$, and solve the resulting quadratic equation for $a$:
\begin{align}
a_i^* &\in \frac{- (r_i^{(2)} - \frac{1}{2\mu}) \pm \sqrt{(r_i^{(2)} - \frac{1}{2\mu})^2 - 4 \tilde{r}_i^{(1)} \tilde{r}_i^{(3)}}}{2 \tilde{r}_i^{(1)}} .
\end{align}
We select larger of the two possible values in range $[a^l, a^u)$ as $a_i^*$.
Finally, we compute $a^* = \min_i a_i^*$.

\subsection{ADMM Optimization Algorithm}

We solve \eqref{eq:obj} via an ADMM framework \cite{parikh13}. 
We introduce auxiliary variable $\z = \C \x$, so that \eqref{eq:obj} can be written as
\begin{align}
\min_{\x,\z} &~ \|\y - \x\|^2_2 + \mu \|\z\|_1 - \mu \, \x^\top \L^p_a(\x') \x
\nonumber \\
\mbox{s.t.} &~ \z = \C \x .
\end{align}
We write the optimization into an unconstrained form via the augmented Lagrangian method \cite{boyd04}:
\begin{align}
\min_{\x,\z} &~ \|\y - \x\|^2_2 + \mu \|\z\|_1 - \mu \, \x^\top \L^p_a(\x') \x
\nonumber \\
&~ + \bxi^\top (\z - \C \x) + \frac{\rho}{2} \left\| \z - \C \x \right\|^2_2
\label{eq:ADMM_obj}
\end{align}
where $\bxi \in \mathbb{R}^M$ is a Lagrange multiplier vector, and $\rho \in \mathbb{R}^+$ is an ADMM parameter.

To solve \eqref{eq:ADMM_obj}, we minimize $\x$ and $\z$ alternately while holding the other variable fixed, until solution convergence.
For notation simplicity, we write $\L^p_a = \L^p_a(\x')$ in the sequel. 

\subsubsection{Optimizing $\x^{t+1}$}

Given fixed $\z^t$ and $\bxi^t$, optimizing $\x$ in \eqref{eq:ADMM_obj} becomes

\vspace{-0.05in}
\begin{small}
\begin{align}
\min_{\x} &~ \|\y - \x\|^2_2 - \mu \, \x^\top \L^p_a \x
+ (\bxi^t)^\top (\z^t - \C \x) + \frac{\rho}{2} \left\| \z^t - \C \x \right\|^2_2
\label{eq:ADMM_obj1}
\end{align}
\end{small}\noindent
which is convex and quadratic.
Thus, the optimal $\x^{t+1}$ can be obtained by solving a linear system:
\begin{align}
\left(2 \I - 2 \mu \L^p_a + \rho \C^\top \C \right) \x^{t+1} = 2 \y + \rho \C^\top \z^t + \C^\top \bxi^t .
\label{eq:linSys}
\end{align}
Given that the coefficient matrix on the left-hand side is sparse, symmetric and PD, $\x^{t+1}$ can be obtained via \textit{conjugate gradient} (CG) \cite{shewchuk94} in linear time.

\subsubsection{Optimizing $\z^{t+1}$}

Given fixed $\x^{t+1}$ and $\bxi^t$, optimizing $\z$ in \eqref{eq:ADMM_obj} becomes

\vspace{-0.1in}
\begin{small}
\begin{align}
\min_{\z} &~ \mu \|\z\|_1 + (\bxi^t)^\top (\z - \C \x^{t+1}) + \frac{\rho}{2} \left\| \z - \C \x^{t+1} \right\|^2_2 .
\label{eq:ADMM_obj2}
\end{align}
\end{small}\noindent
\eqref{eq:ADMM_obj2} is a sum of a convex smooth function $f(\z) = (\bxi^t)^\top (\z - \C \x^{t+1}) + \frac{\rho}{2} \left\| \z - \C \x^{t+1} \right\|^2_2$ and convex non-smooth function $g(\z) = \mu \|\z\|_1$. 
We can thus solve \eqref{eq:ADMM_obj2} via \textit{proximal gradient descent} (PGD) \cite{parikh13}.
Specifically, for each \textit{inner} iteration $\tau$, we compute solution $\z^{\tau+1}$ from previous solution $\z^\tau$ as
\begin{align}
\z^{\tau+1} = \text{prox}_{\lambda g}(\z^\tau - \gamma \nabla f(\z^\tau))
\end{align}
where $\gamma > 0$ is a step size. 
The gradient $\nabla f(\z)$ of $f(\z)$ is 
\begin{align}
\nabla f(\z) = \bxi^t + \rho (\z - \C \x^{t+1}) ,
\end{align}
and the proximal mapping for $g(\z)$ is
\begin{align}
\text{prox}_{\lambda g}(\z) = \min_\u \mu \|\u\|_1 + \frac{1}{2 \lambda} \|\u - \z \|^2_2 
\label{eq:prox}
\end{align}
where $\lambda > 0$ is a proximal mapping parameter.
It is known that the solution to \eqref{eq:prox} is a term-by-term soft-thresholding operation \cite{parikh13}:
\begin{align}
\text{prox}_{\lambda g}(z_i) = \text{sign}(z_i) \cdot \max \left( |z_i| - \lambda \mu, 0 \right) .
\label{eq:soft}
\end{align}

\subsubsection{Updating $\bxi^{t+1}$}

Given fixed $\x^{t+1}$ and $\z^{t+1}$, Lagrange multiplier $\bxi^{t+1}$ can be updated in the conventional manner in ADMM \cite{parikh13}:
\begin{align}
\bxi^{t+1} = \bxi^{t} + \rho \left( \z^{t+1} - \C \x^{t+1} \right) .
\end{align}

\section{Algorithm Unrolling}
\label{sec:unroll}
We next unroll each iteration of our ADMM algorithm into a neural layer to compose a feed-forward network for data-driven parameter learning. 
In addition, we learn a \textit{similarity graph} $\cG$ from data, so that incidence matrix $\C$ specifying a positive graph $\cG$ in \eqref{eq:obj} is properly defined.

\subsection{Similarity Graph Learning}

We define a positive graph $\cG$ with edge weights $w_{i,j}$'s that are exponential functions of \textit{Mahalanobis distance} $d_{i,j}$'s:
\begin{align}
w_{i,j} &= \exp \left( - d_{i,j} \right), 
\\
d_{i,j} &= (\f_i - \f_j)^\top \M (\f_i - \f_j) 
\end{align}
where $\M \in \mathbb{R}^{K \times K}$ is a PSD \textit{metric matrix}, and $\f_i \in \mathbf{R}^K$ is a $K$-dimensional feature vector representing pixel $i$.
We compute $\f_i$ via a shallow CNN from an input embedding $\e_i \in \mathbb{R}^E$ that is an image patch centered at pixel $i$: 
$\f_i = \text{CNN}(\e_i)$.

\subsection{Unrolled Network Architecture}

\begin{figure}[htbp]
\noindent \includegraphics[width=3.4in]{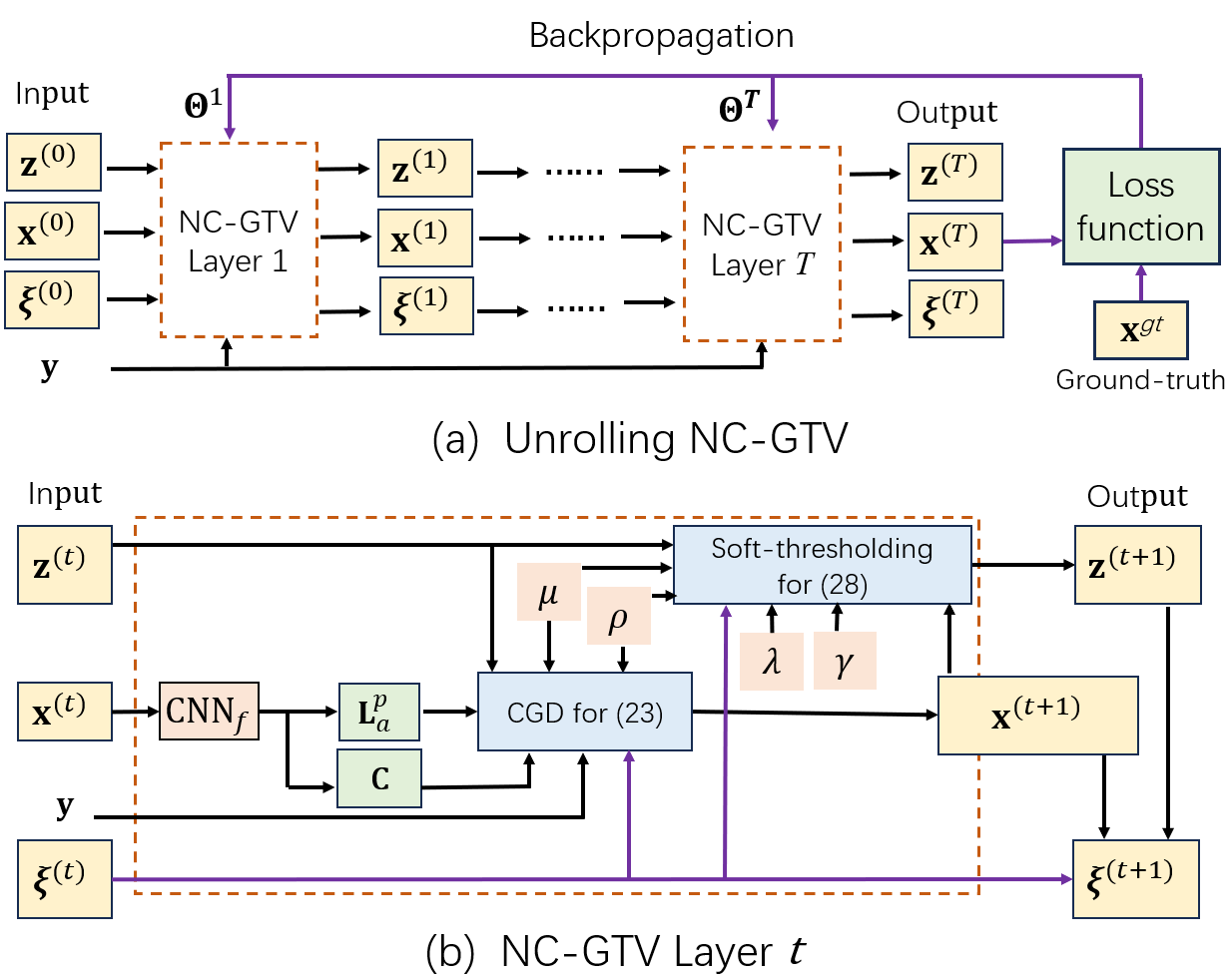}
\vspace{-0.2in}
  \caption{Overview of the proposed architecture. (a) Unrolled NC-GTV consists of multiple NC-GTV Layers, (b) Block diagram of NC-GTV Layer.}
  \label{fig:unrolledNetwork}
\end{figure}

We unroll our ADMM algorithm into $T$ NC-GTV layers.
Fig.\;\ref{fig:unrolledNetwork} shows the unrolled network architecture. 
Each NC-GTV layer has the same structure and parameters. 
To learn edge weights from data to define $\C$ and $\L_{a}^{p}$, we employ a shallow CNN to learn feature vectors $\f_i$ per pixel $i$, and then compute edge weights.
Parameters $\mu$, $\rho$, $\gamma$ and $\lambda$ in \eqref{eq:ADMM_obj1} and \eqref{eq:ADMM_obj2} are trainable parameters in the unrolled network.

Similarly done in \cite{vu21}, the input to the algorithm is 
$\mathbf{z}^{(0)}=\mathbf{C}\mathbf{y}$ and $\mathbf{\bxi}^{(0)}=\mathbf{0}$.
After $T$ layers, we obtain the reconstruction signal $\mathbf{x}^{(T)}$. 
The $t$-th NC-GTV layer uses the output of previous layer, $\mathbf{z}^{(t-1)}$, $\x^{(t-1)}$ and $\mathbf{\bxi}^{(t-1)}$ and a set of learned parameters 
$\mathbf{\Theta}^{t}=\{\rho^{(t)},\mu^{(t)},\lambda^{(t)}, \gamma^{(t)}, 
\mathbf{\theta}_{f}^{(t)}\}$ as input, where $\mathbf{\theta}_f^{(t)}$ are the parameters of $\mathrm{CNN}_{f}$.
To train our unrolled network, we evaluate the \textit{mean squared error} (MSE) between the recovered image $\mathbf{x}^{(T)}$ and the ground-truth image $\mathbf{x}^{gt}$ for $K$ training patches to update parameters $\{\mathbf{\Theta}^1, \cdots, \mathbf{\Theta}^T\}$.

\section{Experiments}
\label{sec:results}
\vspace{-0.05in}
\subsection{Experimental Setup}

\vspace{-0.05in}
We set the number of NC-GTV layers to $T=2$. 
$\mathrm{CNN}_f$ consists of $4$ convolution layers. 
The first layer has $3$ input channels and $32$ output channels, while  
the last layer has $32$ and $3$.
Except for the last layer, we used $\operatorname{ReLU}(\cdot)$ after every convolutional layer.
As done in \cite{vu21}, we evaluated our model on the RENOIR dataset and added Gaussian noise to the clean images for experiments. 
Images were divided into patches of size 36 $\times$ 36. 
We trained our model for $200$ epochs in each experiment using \textit{stochastic gradient descent} (SGD). 
The batch size and learning rate were set to $10$ and $1 \times 10^{-4}$. 
Our model was implemented in PyTorch and trained on an Nvidia L40 GPU. 
We compared our model with representative image denoising schemes: model-based CBM3D\,\cite{dabov07}, and DL-based CDnCNN\,\cite{DnCNN}, DGLR\,\cite{zeng19} and DGTV\,\cite{vu21}. 
We evaluated the performance using peak signal-to-noise ratio (PSNR) and structural similarity index measure (SSIM) as metrics. 

\subsection{Experimental Results}
We first trained and tested on additive white Gaussian noise (AWGN) with $\sigma=30$. 
Results are shown in Table\;\ref{table_result}. 
We observe that learning-based methods generally outperformed model-based CBM3D. 
Compared to CDnCNN, our model achieved comparable performance, while using only 18\% of network parameters. 
Our method performed better than previous graph-based schemes DGTV and DGLR. 

\begin{figure}[htbp]
\noindent 
\centering
\includegraphics[width=3.3in]{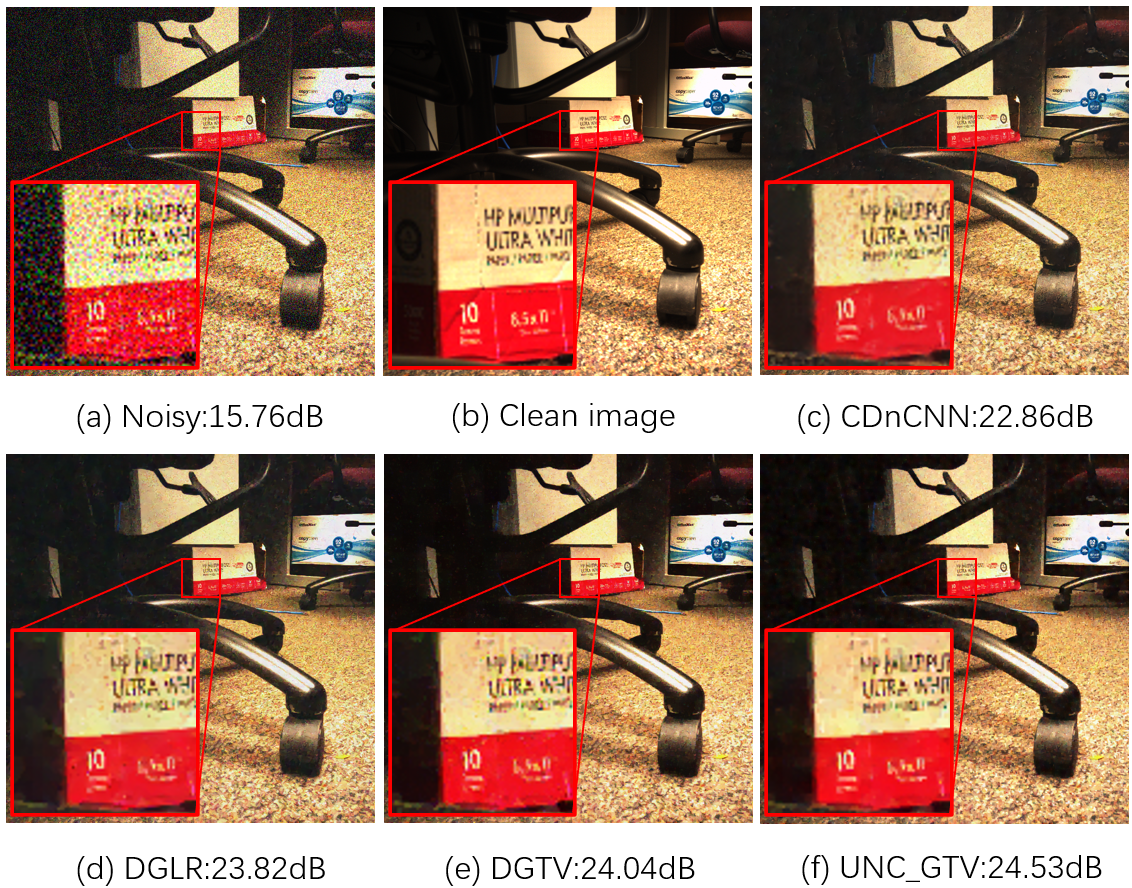}
\vspace{-0.1in}
  \caption{Image examples under a noise variance mismatch scenario: Training on $\sigma=30$ and testing on $\sigma=50$.}
  \label{fig:result_sig50}
  \vspace{-0.1in}
\end{figure}

To demonstrate robustness against noise variance mismatch, we trained all models on AWGN with $\sigma=30$ and tested them with $\sigma=50$. 
As shown in Table\;\ref{table_result}, NC-GTV  achieved the highest PSNR and SSIM. Fig.\;\ref{fig:result_sig50} shows example images of various models. 

\begin{table}[ht]
    \centering
    \caption{Number of trainable parameters, average PSNR and SSIM in AWGN denoising and noise variance mismatch.}
    \vspace{-0.1in}
    \footnotesize
    \begin{tabular}{|c|c|c|c|c|c|}
        \hline
        \multirow{3}{*}{Method }& \multirow{3}{*}{\#Parameters} & \multicolumn{2}{c|}{\parbox[t]{1.4cm}{\centering{$\sigma_{\text{train}}= 30$\\
        $\sigma_{\text{test}}=30$}}}& \multicolumn{2}{c|}{\parbox[t]{1.4cm}{\centering{$\sigma_{\text{train}}= 30$\\        $\sigma_{\text{test}}=50$\bigstrut[b]} }}  
        \\        
        \cline{3-6}
        &   & PSNR & SSIM & PSNR & SSIM \\
        \hline
        CBM3D & N/A & 29.11 & 0.766 & N/A & N/A \\
        \hline
        CDnCNN & 0.56M & \textbf{29.61} & 0.769 & 23.88 & 0.538 \\
        \hline
        DeepGLR & 0.93M & 29.43 & 0.746 & 24.09 & 0.538 \\
        \hline
        DeepGTV & 0.21M & 29.36 & \textbf{0.772} & 24.91 & 0.572 \\
        \hline
        NC-GTV & \textbf{0.10M} & 29.60 & 0.770 & \textbf{25.43} & \textbf{0.593} \\
        \hline
    \end{tabular}
    \label{table_result}
    \vspace{-0.1in}
\end{table}


\section{Conclusion}
\label{sec:conclude}
Instead of convex total variation (TV) regularization term that underestimates signal discontinuities, we propose non-convex graph total variation (NC-GTV): combine a non-convex regularization term defined by a graph variant of the Huber function with an $\ell_2$-norm fidelity term to compose an overall convex objective, thus mitigating extraneous local minima. 
We show that our graph Huber function can be understood as a Moreau envelope. 
We design an iterative optimization algorithm based on alternating direction method of multipliers (ADMM), and unroll it into a lightweight feed-forward network for data-driven parameter learning.
Experimental results show that our method has comparable image denoising performance as representative deep learning models, while using a fraction of network parameters. 

\pagebreak

\begin{small}
\bibliographystyle{IEEEbib}
\bibliography{ref2}

\begin{thebibliography}{10}

\bibitem{milanfar13}
P.~Milanfar,
\newblock ``A tour of modern image filtering: New insights and methods, both
  practical and theoretical,''
\newblock {\em IEEE Signal Processing Magazine}, vol. 30, no. 1, pp. 106--128,
  2013.

\bibitem{bai19}
Y.~Bai, G.~Cheung, X.~Liu, and W.~Gao,
\newblock ``Graph-based blind image deblurring from a single photograph,''
\newblock {\em IEEE Transactions on Image Processing}, vol. 28, no.3, pp.
  1404--1418, 2019.

\bibitem{chan17}
S.~H. Chan, X.~Wang, and O.~A. Elgendy,
\newblock ``Plug-and-play {ADMM} for image restoration: Fixed-point convergence
  and applications,''
\newblock {\em IEEE Transactions on Computational Imaging}, vol. 3, no. 1, pp.
  84--98, 2017.

\bibitem{chan24}
Stanley~H. Chan,
\newblock ``Tutorial on diffusion models for imaging and vision,'' arXiv
  preprint arXiv:2403.18103, March 2024,
\newblock Last revised January 2025.

\bibitem{DnCNN}
K.~Zhang, W.~Zuo, Y.~Chen, D.~Meng, and L.~Zhang,
\newblock ``Beyond a {Gaussian} denoiser: Residual learning of deep {CNN} for
  image denoising,''
\newblock {\em IEEE Transactions on Image Processing}, vol. 26, no. 7, pp.
  3142--3155, 2017.

\bibitem{monga21}
V.~Monga, Y.~Li, and Y.~C. Eldar,
\newblock ``Algorithm unrolling: Interpretable, efficient deep learning for
  signal and image processing,''
\newblock {\em IEEE Signal Processing Magazine}, vol. 38, no. 2, pp. 18--44,
  2021.

\bibitem{ortega18ieee}
A.~Ortega, P.~Frossard, J.~Kovacevic, J.~M.~F. Moura, and P.~Vandergheynst,
\newblock ``Graph signal processing: Overview, challenges, and applications,''
\newblock in {\em Proceedings of the {IEEE}}, May 2018, vol. 106, no.5, pp.
  808--828.

\bibitem{cheung18}
G.~Cheung, E.~Magli, Y.~Tanaka, and M.~Ng,
\newblock ``Graph spectral image processing,''
\newblock in {\em Proceedings of the {IEEE}}, 2018, vol. 106, no.5, pp.
  907--930.

\bibitem{strong03}
D.~Strong and T.~Chan,
\newblock ``Edge-preserving and scale-dependent properties of total variation
  regularization,''
\newblock {\em Inverse Problems}, vol. 19, no. 6, pp. S165, nov 2003.

\bibitem{Selesnick2020}
I.~Selesnick, A.~Lanza, S.~Morigi, and F.~Sgallari,
\newblock ``Non-convex total variation regularization for convex denoising of
  signals,''
\newblock {\em Journal of Mathematical Imaging and Vision}, vol. 62, no. 6, pp.
  825--841, Jul 2020.

\bibitem{selesnick15}
I.~Selesnick, A.~Parekh, and İ. Bayram,
\newblock ``Convex {1-D} total variation denoising with non-convex
  regularization,''
\newblock {\em IEEE Signal Processing Letters}, vol. 22, no. 2, pp. 141--144,
  2015.

\bibitem{lanza2019sparsity}
Alessandro Lanza, Serena Morigi, Ivan~W. Selesnick, and Fiorella Sgallari,
\newblock ``Sparsity-inducing nonconvex nonseparable regularization for convex
  image processing,''
\newblock {\em SIAM Journal on Imaging Sciences}, vol. 12, no. 2, pp.
  1099--1134, 2019.

\bibitem{huber64}
P.~J. Huber,
\newblock ``Robust estimation of a location parameter,''
\newblock {\em Ann. Math. Statist.}, vol. 35, no. 1, pp. 73--101, 1964.

\bibitem{heinz17}
H.~B. Heinz and P.~L. Combettes,
\newblock {\em Convex Analysis and Monotone Operator Theory in Hilbert Spaces},
\newblock Springer, 2017.

\bibitem{varga04}
R.~S. Varga,
\newblock {\em {Gershgorin} and his circles},
\newblock Springer, 2004.

\bibitem{parikh13}
N.~Parikh and S.~Boyd,
\newblock ``Proximal algorithms,''
\newblock in {\em Foundations and Trends in Optimization}, 2013, vol. 1, no.3,
  pp. 123--231.

\bibitem{dabov07}
K.~Dabov, A.~Foi, V.~Katkovnik, and K.~Egiazarian,
\newblock ``Image denoising by sparse {3-D} transform-domain collaborative
  filtering,''
\newblock {\em IEEE Transactions on Image Processing}, vol. 16, no. 8, pp.
  2080--2095, 2007.

\bibitem{pang17}
J.~Pang and G.~Cheung,
\newblock ``Graph {Laplacian} regularization for inverse imaging: Analysis in
  the continuous domain,''
\newblock in {\em IEEE Transactions on Image Processing}, April 2017, vol. 26,
  no.4, pp. 1770--1785.

\bibitem{liu17}
X.~Liu, G.~Cheung, X.~Wu, and D.~Zhao,
\newblock ``Random walk graph {Laplacian} based smoothness prior for soft
  decoding of {JPEG} images,''
\newblock {\em IEEE Transactions on Image Processing}, vol. 26, no.2, pp.
  509--524, February 2017.

\bibitem{elmoataz08}
Abderrahim Elmoataz, Olivier Lezoray, and SÉbastien Bougleux,
\newblock ``Nonlocal discrete regularization on weighted graphs: A framework
  for image and manifold processing,''
\newblock {\em IEEE Transactions on Image Processing}, vol. 17, no. 7, pp.
  1047--1060, 2008.

\bibitem{couprie13}
Camille Couprie, Leo Grady, Laurent Najman, Jean-Christophe Pesquet, and Hugues
  Talbot,
\newblock ``Dual constrained {TV}-based regularization on graphs,''
\newblock {\em SIAM Journal on Imaging Sciences}, vol. 6, no. 3, pp.
  1246--1273, 2013.

\bibitem{clr}
T.~H. Cormen, C.~E. Leiserson, and R.~L. Rivest,
\newblock {\em Introduction to Algorithms},
\newblock McGraw Hill, 1986.

\bibitem{boyd04}
S.~Boyd and L.~Vandenberghe,
\newblock {\em Convex Optimization},
\newblock Cambridge, 2004.

\bibitem{shewchuk94}
J.~R Shewchuk,
\newblock ``An introduction to the conjugate gradient method without the
  agonizing pain,''
\newblock Tech. {R}ep., USA, 1994.

\bibitem{vu21}
Huy Vu, Gene Cheung, and Yonina~C. Eldar,
\newblock ``Unrolling of deep graph total variation for image denoising,''
\newblock in {\em ICASSP 2021 - 2021 IEEE International Conference on
  Acoustics, Speech and Signal Processing (ICASSP)}, 2021, pp. 2050--2054.

\bibitem{zeng19}
Jin Zeng, Jiahao Pang, Wenxiu Sun, and Gene Cheung,
\newblock ``Deep graph {Laplacian} regularization for robust denoising of real
  images,''
\newblock in {\em 2019 IEEE/CVF Conference on Computer Vision and Pattern
  Recognition Workshops (CVPRW)}, 2019, pp. 1759--1768.

\end{thebibliography}
\end{small}

\end{document}